\documentclass[preprint2,twocolumn,times,tighten]{aastex631}

\usepackage{graphicx}	
\usepackage{amsmath}	
\usepackage{multirow}
\let\tablenum\relax
\usepackage{siunitx}

\begin{document}

\title{Characterizing 3D Magnetic Fields and Turbulence in H I Clouds}

\author[0000-0002-8455-0805]{Yue Hu*}
\affiliation{Institute for Advanced Study, 1 Einstein Drive, Princeton, NJ 08540, USA}


\email{yuehu@ias.edu, *NASA Hubble Fellow}



\begin{abstract}
3D Galactic magnetic fields are critical for understanding the interstellar medium, Galactic foreground polarization, and the propagation of ultra-high-energy cosmic rays. Leveraging recent theoretical insights into anisotropic magnetohydrodynamic (MHD) turbulence, we introduce a deep learning framework to predict the full 3D magnetic field structure—including the plane-of-sky (POS) position angle, line-of-sight (LOS) inclination, magnetic field strength, sonic Mach number ($M_s$), and Alfv\'en Mach number ($M_A$)—from spectroscopic H~I observations. The deep learning model is trained on synthetic H~I emission data generated from multiphase 3D MHD simulations. We then apply the trained model to observational data from the Commensal Radio Astronomy FAST Survey, presenting maps of 3D magnetic field orientation, magnetic field strength, $M_s$, and $M_A$ for two H~I clouds—a low-velocity cloud (LVC) and an intermediate-velocity cloud (IVC)—which overlap in the POS yet reside at different LOS distances. The deep-learning–predicted POS magnetic field position angles align closely with those determined using the velocity gradient technique, whose integrated results are consistent with independent measurements from Planck 353~GHz polarization data. This study demonstrates the potential of deep learning approaches as powerful tools for modeling the 3D distributions of 3D Galactic magnetic fields and turbulence properties throughout the Galaxy.
\end{abstract}


\keywords{Interstellar medium (847) --- Neutral hydrogen clouds (1099) --- Interstellar magnetic fields (845) --- Magnetohydrodynamical simulations (1966) --- Deep learning (1938)}


\section{Introduction} 
\label{sec:intro}
\begin{figure}
\centering
\includegraphics[width=0.9\linewidth]{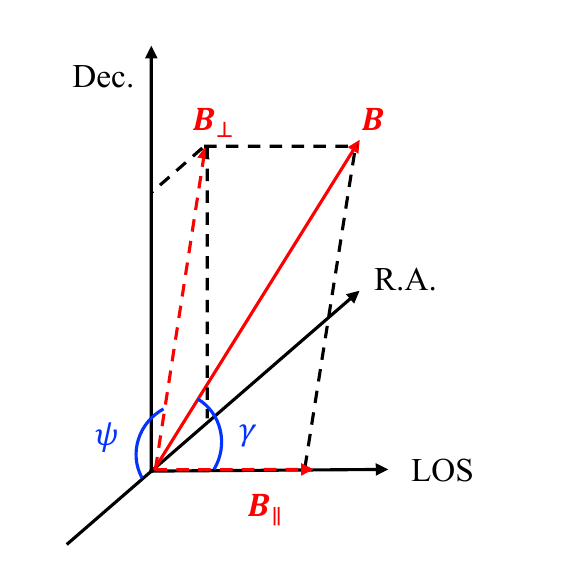}
        \caption{Definition of the 3D magnetic field $\pmb{B}$. $\pmb{B}_\bot$ is the magnetic field projected on the POS and $\pmb{B}_\parallel$ is the LOS component. $\psi$ is $\pmb{B}_\bot$’s position angle relative on the POS. $\gamma$ is $\pmb{B}$'s inclination angle with respect to the LOS. }
    \label{fig:B}
\end{figure}

Magnetic fields pervade the interstellar medium (ISM) and play critical roles in governing fundamental astrophysical processes, including star formation, gas dynamics, and cosmic ray propagation \citep{2012ARA&A..50...29C,2015ARA&A..53..501A,2022ApJ...941...92H,2020ApJ...901..162H,2019NatAs...3..776H,2020ApJ...894...63X,2022MNRAS.510.4952L,2025ApJS..276...15P}. Accurate characterization of the 3D magnetic field (see Fig.~\ref{fig:B} for the definition) — including its plane-of-the-sky (POS) orientation, inclination angle to the line-of-sight (LOS), strength, and associated turbulence parameters—remains an essential yet challenging task \citep{2023MNRAS.519.3736H,2024ApJ...965..183H,2024MNRAS.52711240H}. In particular, precise modeling of the Galactic magnetic field is essential for accurate removal of the Galactic polarized foreground emission, a critical step for separating cosmological signals, such as the cosmic microwave background (CMB) polarization \citep{2014PhRvL.112x1101B,2016A&A...586A.133P}.  Furthermore, understanding the 3D structure of Galactic magnetic fields is indispensable for accurately tracing the propagation trajectories of ultra-high-energy cosmic rays (UHECRs), for instance, the Amaterasu \citep{2019JCAP...05..004F,2023Sci...382..903T,2025arXiv250215876U}.

Despite their importance, directly probing the 3D magnetic fields and characterizing the turbulence parameters - sonic Mach number $M_s$ and Alfv\'en Mach number $M_A$ - remain an outstanding observational challenge. Traditional observational techniques provide either POS orientation through polarized dust and synchrotron emission \citep{2007JQSRT.106..225L,2020A&A...641A..11P,2020A&A...641A...4P} or LOS components via Zeeman splitting and Faraday rotation \citep{2012ARA&A..50...29C,2012A&A...542A..93O,2019A&A...632A..68T}. However, combining these separate measurements into a 3D magnetic field vector is non-trivial, particularly in the presence of multiphase and turbulent ISM conditions.

An alternative approach has emerged from recent theoretical and numerical studies of anisotropic magnetohydrodynamic (MHD) turbulence, demonstrating that turbulent eddies preferentially elongate along local magnetic fields \citep{GS95,LV99} and their anisotropy degree is correlated with magnetic field strength \citep{2024ApJ...974..237L}. This anisotropy is imprinted in spectroscopic observations, proving an independent way to study the magnetic fields \citep{2000ApJ...537..720L,2016MNRAS.461.1227K,2021ApJ...910..161Y,2023MNRAS.524.2994H}. Leveraging these theoretical insights, deep learning techniques have shown remarkable promise in extracting magnetic field and turbulence information embedded in spectroscopic observations \citep{2024MNRAS.52711240H,2024arXiv241111157S,2025ApJ...980...52X}. Particularly, \cite{2021ApJ...915...67H,2024MNRAS.52711240H} explained the corresponding physics and introduced a convolutional neural network framework trained on synthetic molecular spectral line observations, reconstructing the 3D magnetic field in the molecular cloud L1478.

In this paper, we extend our previous deep-learning-based approach beyond molecular clouds to the more diffuse and multiphase atomic hydrogen (H I) environments. Such diffuse H I clouds present distinct challenges due to their thermal instability and complex multiphase turbulence properties \citep{1977ApJ...218..148M,1995ApJ...443..152W,2000ApJ...540..271V,2003ApJ...587..278W}. To address these challenges, we have refined and implemented a new deep learning architecture that simultaneously predicts 3D magnetic field orientation, magnetic field strength, $M_s$, and $M_A$. We train our deep learning model using synthetic H I emission data generated from 3D MHD simulations of multiphase ISM.

As a direct observational validation, we apply our trained deep learning model to H I data from the Five-hundred-meter Aperture Spherical radio Telescope (FAST) Commensal Radio Astronomy FAST Survey (CRAFTS; \citealt{2018IMMag..19..112L}) H I observations, focusing specifically on two HI clouds - a low velocity cloud (LVC) and an intermediate velocity cloud (IVC)- around the Monoceros region. The two overlap in the POS but with different LOS distances. This application not only provides the first 3D magnetic field characterization for diffuse H I clouds, but also shows the advantages of recovering the 3D magnetic field's variation along the LOS.


This paper is structured as follows: In \S~\ref{sec:theory}, we briefly review the theoretical basis of anisotropic MHD turbulence, its relationship to magnetic field structure and turbulence properties, and the anisotropy's imprints in spectroscopic observation. \S~\ref{sec:method} describes our deep learning architecture, training procedure, and synthetic data generation from numerical simulations. In \S~\ref{sec:results}, we present numerical validation of the deep learning model, quantifying its performance across different physical conditions. We further apply the model to predict the 3D magnetic field in the LVC and IVC using the CRAFTS H I survey data. In \S~\ref{sec:discussion}, we discuss the potential synergies with other methods, and prospects of the deep-learning-based approach. We summarize the key conclusions of our study in \S~\ref{sec:conclusion}.

\section{Theoretical consideration}
\label{sec:theory}
\subsection{MHD turbulence's anisotropy reveals 3D magnetic field orientation and magnetization}
The physical basis for tracing the 3D magnetic fields using spectroscopic observations hinges on the anisotropy of MHD turbulence (see \citealt{2024MNRAS.52711240H} for details). Here we briefly review the important concepts.

In the turbulent and magnetized ISM, the cascade of turbulent energy is not isotropic; rather, turbulent eddies are preferentially elongated along the magnetic field lines as a direct consequence of the critical balance condition proposed by \cite{GS95} and \cite{LV99}. This anisotropy has been confirmed by isothermal and multiphase numerical simulations \citep{2000ApJ...539..273C, 2001ApJ...554.1175M, 2003MNRAS.345..325C, 2010ApJ...720..742K, HXL21, 2024MNRAS.527.3945H,Hu2025} and in situ solar wind measurements \citep{2016ApJ...816...15W, 2020FrASS...7...83M, 2021ApJ...915L...8D, 2024NatAs...8..725Z}. Under this paradigm, the eddy turnover time for motions perpendicular to the field becomes comparable to the Alfv\'en wave crossing time along the field. As a result, the turbulent eddies have three important properties:
\begin{enumerate}
    \item The elongation direction reveals the local magnetic field orientation.
    \item The ratio of scales parallel and perpendicular to the magnetic field (and, hence, the degree of anisotropy) directly reflects the local level of magnetization $M_A^{-1}$. A strongly magnetized medium shows a large anisotropy degree.
    \item The curvature is correlated with the magnetization. The stronger the magnetization, the less curved elongation of the eddies.
\end{enumerate}

These properties are imprinted in spectroscopic observations due to the velocity caustics effect \citep{2000ApJ...537..720L}. The effect means that the observed intensity distribution of a given spectral channel is determined by both the density of emitters and their velocity distribution along the LOS. Provided that the channel width (i.e., velocity resolution) is smaller than the turbulent velocity dispersion, the intensity fluctuations in the thin channel become dominated by velocity caustics. These velocity caustics retain statistical properties of the anisotropic turbulence; the elongated structures, aligned with the POS magnetic field, become readily apparent, providing a crucial feature to trace the POS magnetic field orientation \citep{2018ApJ...853...96L,2021ApJ...910..161Y,2023MNRAS.524.2994H}.

Moreover, the observed anisotropy, as well as the observed curvature, is sensitive not only to the magnetization but also to projection effects introduced by the inclination of the magnetic field relative to the observer’s LOS. The observed anisotropy degree and curvature will change as the magnetic field is inclined close to the LOS. By carefully accounting for these projection effects and the modulation of anisotropy and curvature by the local magnetization, one can extract both the POS magnetic field orientation $\psi$, the magnetic field's inclination angle $\gamma$, and the magnetization $M_A^{-1}$ \citep{2024MNRAS.52711240H}.

\subsection{Shock effects reveal the sonic Mach number}
Spectroscopic observations contain the information of the sonic Mach number $M_s$. A supersonic medium is affected by shocks: the larger $M_s$, the more shocks. Shocks introduce compression to the gas and create small-scale filamentary structures in spectral channels. These small-scale filamentary structures serve as a distinct feature to measure the $M_s$
\citep{2024arXiv241111157S}.

It is important to note that the squared $M_A$ reflects the relative importance of turbulent kinetic energy and magnetic field energy, while the squared $M_s$ characterizes the ratio of turbulent kinetic energy and thermal kinetic energy. Combining two Mach numbers, the magnetic field strength $B$ can be expressed as \citep{2022ApJ...935...77L}:
\begin{equation}
\label{eq.B}
    B=c_s\sqrt{4\pi\rho}M_sM_A^{-1},
\end{equation}
where $c_s$ is the sound speed and $\rho$ is gas mass density. It suggests that spectroscopic observations contain the information of $B$. Instead of measuring the two Mach numbers, \cite{2024arXiv241107080Z} confirmed that machine learning tools can extract $B$ from observations directly.

\begin{figure}
\centering
\includegraphics[width=1.0\linewidth]{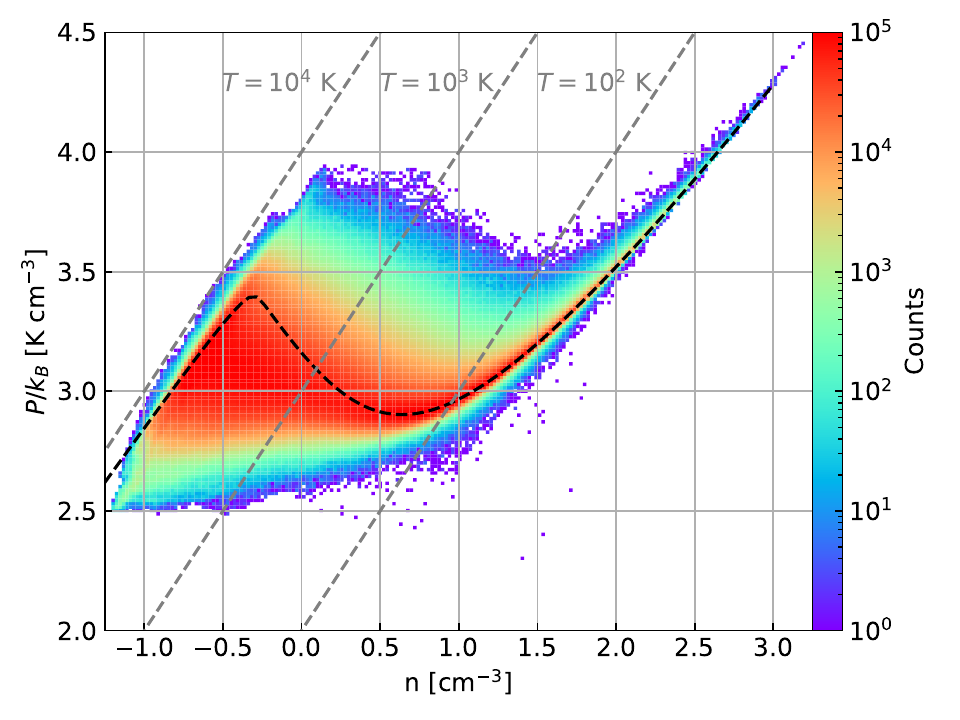}
        \caption{Phase diagrams of gas number density and pressure. The black dashed line represents the thermal equilibrium obtained from $\Gamma = \Lambda$, where $\Gamma$ and $\Lambda$ are the heating and cooling functions, respectively. The simulation with $B\approx \SI{3}{\micro G}$ and $\sigma_v\approx$ 5.00 km s$^{-1}$ is used.}
    \label{fig:phase_diagram}
\end{figure}

\section{Methodology}
\label{sec:method}
\subsection{Numerical simulations}
The 3D multi-phase ISM simulations employed in this study were generated using the AthenaK code \citep{2024arXiv240916053S}. These simulations solve the ideal MHD equations under periodic boundary conditions. The energy equation incorporates atomic line cooling and photoelectric heating processes \citep{2002ApJ...564L..97K}. An example of the simulation's phase diagram is presented in Fig.~\ref{fig:phase_diagram}. Further details regarding the simulation setup and numerical methods are provided in \cite{Hu2025}.

The simulation box size is 100~pc, with turbulence driven solenoidally at a peak wavenumber of 2. The computational domain is discretized on a uniform $512^3$ grid, with numerical dissipation occurring at scales of approximately 10 cells. The initial conditions include a uniform number density field of $n = 3~{\rm cm^{-3}}$ and a uniform magnetic field aligned along the $y$-axis. The simulation cubes were subsequently rotated to align the mean magnetic field inclination with respect to the LOS, or the $z$-axis, at angles of 90$^\circ$, 60$^\circ$, and 30$^\circ$, respectively. We choose magnetic field strength $B \approx1$, 3, and 5 \SI{}{\micro G} based on Zeeman observation \citep{2012ARA&A..50...29C}. 

Three values of velocity dispersion are included: $\sigma_v\approx1.25$, 2.50, and 5.00 km s$^{-1}$. The choice is based on Larson's law and observational constraints from young stars \citep{1981MNRAS.194..809L,2022ApJ...934....7H,2023AAS...24122801H}. Combining these magnetic field strengths, velocity dispersions, and inclination angles yields a total of 27 unique simulation parameter sets, covering mean $M_A$ approximately from 0.2 to 5 and mean $M_s$ approximately from 0.25 to 1. Local values of $M_A$ and $M_s$ may vary with cold neutral medium being typically super-sonic and super-Alfv\'enic. For each parameter set, two snapshots were extracted after the simulations evolved for 100 Myr, ensuring that turbulence had reached a saturated state. Consequently, a total of 54 simulations are used to generate synthetic spectroscopic observations and thin velocity channel $p$ (see \citealt{2023MNRAS.524.2994H} for details) for the neural network training. One more simulation with $B \approx4$ \SI{}{\micro G} and  $\sigma_v\approx2.5$ km s$^{-1}$ is used as the unseen data for testing the neural network. 

\subsection{Observational data}
The H I observations used in this study are drawn from the CRAFTS survey \citep{2018IMMag..19..112L}. In the Monoceros R2 region, the data reveal two distinct velocity components—a LVC and an IVC—that, while overlapping in the POS, are clearly separated in LOS distance. The H I data cube has a beam resolution of approximately 4 arcminutes, and a velocity resolution of 0.2 km s$^{-1}$. The root mean squared noise is around 0.125 K per 0.2 km s$^{-1}$.
\begin{figure*}
\centering
\includegraphics[width=0.99\linewidth]{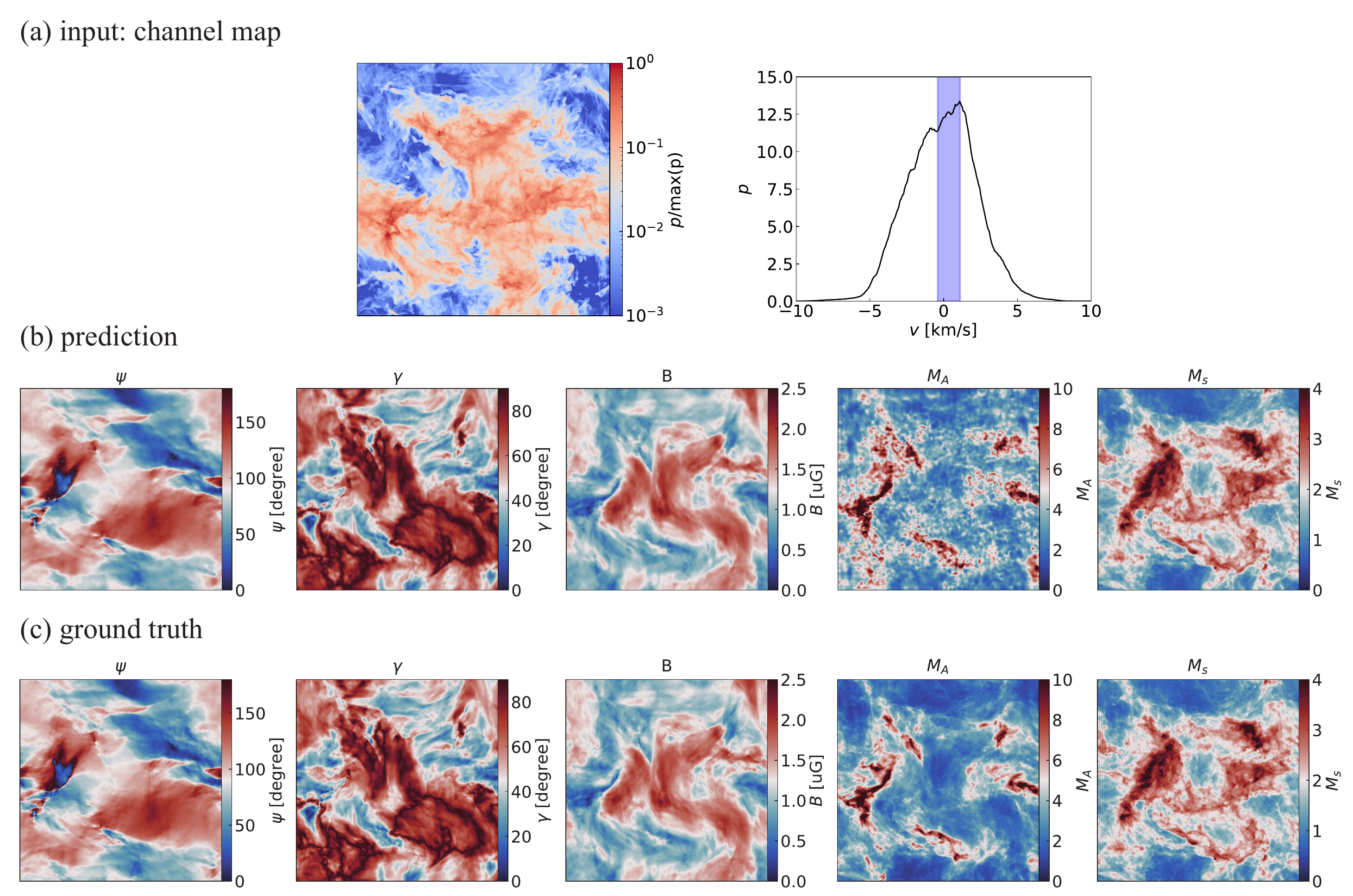}
        \caption{A comparison of the predicted physical quantities and the ground truth. The simulation with mean $B\approx\SI{1}{\micro G}$, $\sigma_v\approx 5.0$ km~s$^{-1}$, and $\gamma\approx60^\circ$ is used as an example. Panel (a): the input velocity channel map $p$ and the corresponding spectrum. The blue shadow area indicates the velocity range used for the channel map integration. The width is 2 km~s$^{-1}$. Panel (b): the predicted POS magnetic field orientation $\psi$, the magnetic field's inclination angle $\gamma$, total magnetic field strength $B$, Alfv\'en Mach number $M_A$, and sonic Mach number $M_s$. Panel (c): the ground truth of the five quantities.}
    \label{fig:v_map}
\end{figure*}

\begin{figure*}
\centering
\includegraphics[width=0.99\linewidth]{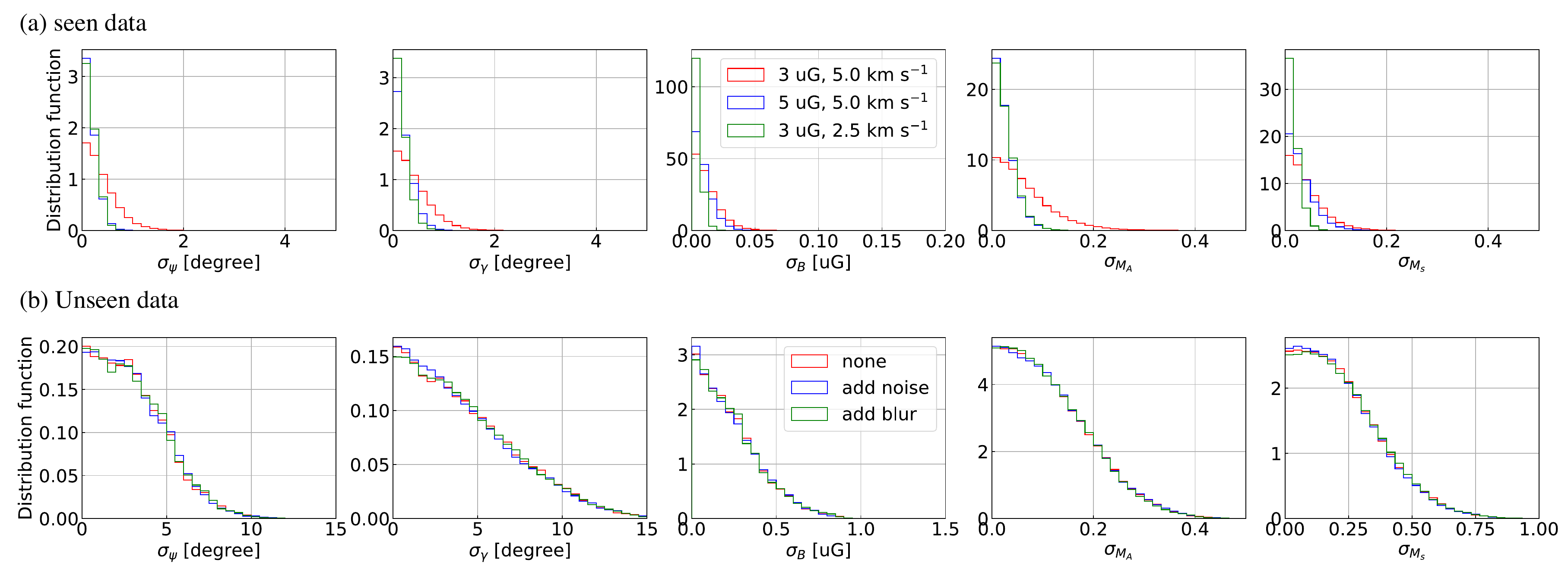}
        \caption{Histograms of the uncertainty $\sigma$ in prediction. $\sigma$ is defined as the absolute difference between the predicted value and the ground truth.  
        Panel (a): $\sigma$ distributions for three simulations that are already used in the training. The three simulations all have a mean $\gamma\approx60^\circ$, but different mean magnetic field strengths ($B\approx\SI{3}{\micro G}$ or $B\approx\SI{5}{\micro G}$) and different velocity dispersions ($\sigma_v\approx 2.5$ km~s$^{-1}$ or $\sigma_v\approx 5.0$ km~s$^{-1}$). Panel (b): $\sigma$ distributions for three simulations that are not used in the training. The three simulations all have a mean $\gamma\approx60^\circ$, $B\approx\SI{4}{\micro G}$, and $\sigma_v\approx 2.5$ km~s$^{-1}$. Two observational effects of noise - reducing the signal-to-noise ratio to 20 - and finite beam size (i.e., Gaussian blur) - lowering the resolution by a factor of 2 - are included.}
    \label{fig:sigma}
\end{figure*}

\subsection{Deep learning model}
We used a deep learning model to infer 3D magnetic fields and turbulence information from spectroscopic data. The model employs a conditional residual neural network (Conditional ResNet) architecture \citep{he2016deep} to predict magnetic field maps from input spectroscopic images. 

\textbf{Neural network architecture:} Our Conditional ResNet integrates conditional information into the prediction process through Feature-wise Linear Modulation (FiLM; \citealt{2017arXiv170907871P}). The network architecture consists of three primary components: (1) Condition Embedding Network \citep{2018arXiv180507544K}: A fully connected neural network transforms input conditioning information into a conditioning vector, which modulates the internal feature representations of the network. (2) Encoder-Decoder Structure \citep{2021arXiv211015253A}: Input spectroscopic images are processed through convolutional layers with instance normalization and reflection padding, followed by downsampling and subsequent upsampling via transposed convolutions. This structure effectively captures spatial features at multiple scales. (3) Conditional Residual Blocks \citep{he2016deep}: Each residual block incorporates FiLM conditioning, dynamically modulating intermediate feature maps based on the embedded conditioning vector. This allows the network to adapt its internal representations to varying input conditions, significantly enhancing prediction accuracy.

\textbf{Training strategy:} The training and evaluation datasets consist of paired spectroscopic images and corresponding output maps ($\psi$, $\gamma$, $B$, $M_A$, or $M_s$). Details of generating those maps are given in \citet{2024MNRAS.52711240H,2024arXiv241111157S}. All input spectroscopic images are individually normalized, so the prediction is only based on morphology. We implemented rotational variations, Gaussian noise, and Gaussian blurring to simulate observational uncertainties and instrumental effects, respectively. The model is trained with 2,000 epochs using the mean squared error loss function, optimized via the Adam optimizer with a learning rate of $2\times10^{-4}$ and momentum parameters (betas) set to 0.5 and 0.999.  To mitigate overfitting, we monitor the validation loss for early stopping and evaluate the peak signal-to-noise ratio (PSNR) on a held-out test set every 50 epochs. The final model corresponds to the checkpoint with the highest PSNR on untransformed test data. Although not implemented in this work, additional regularization techniques such as L2 weight decay and dropout in the conditional residual blocks could further reduce overfitting and may be considered in future extensions.

\section{Results} 
\label{sec:results}
\subsection{A comparison of prediction and ground truth}
Fig.~\ref{fig:v_map} compares the predicted physical parameters - $\psi$, $\gamma$, $B$, $M_A$, and $M_s$ - with the corresponding ground-truth distributions. In this example, the simulation has a mean magnetic field of $B\approx\SI{1}{\micro G}$, a velocity dispersion of $\sigma_v\approx5.0$ km~s$^{-1}$, and a mean inclination angle of $\gamma\approx60^\circ$. The input is a normalized thin velocity channel $p$ with a channel width 2 km~s$^{-1}$, which exhibits highly filamentary intensity structures. The output maps reveal significant fluctuations:  $\psi$, ranging from 0 to 180$^\circ$, and $\gamma$ spanning from 0 to 90$^\circ$ indicate substantial variations in the magnetic field geometry. Moreover, regions with highly super-Alfv\'enic conditions ($M_A\gg1$) and supersonic motions ($M_s>1$) are apparent, typically corresponding to the cold, dense phase.
\begin{figure*}
\centering
\includegraphics[width=0.99\linewidth]{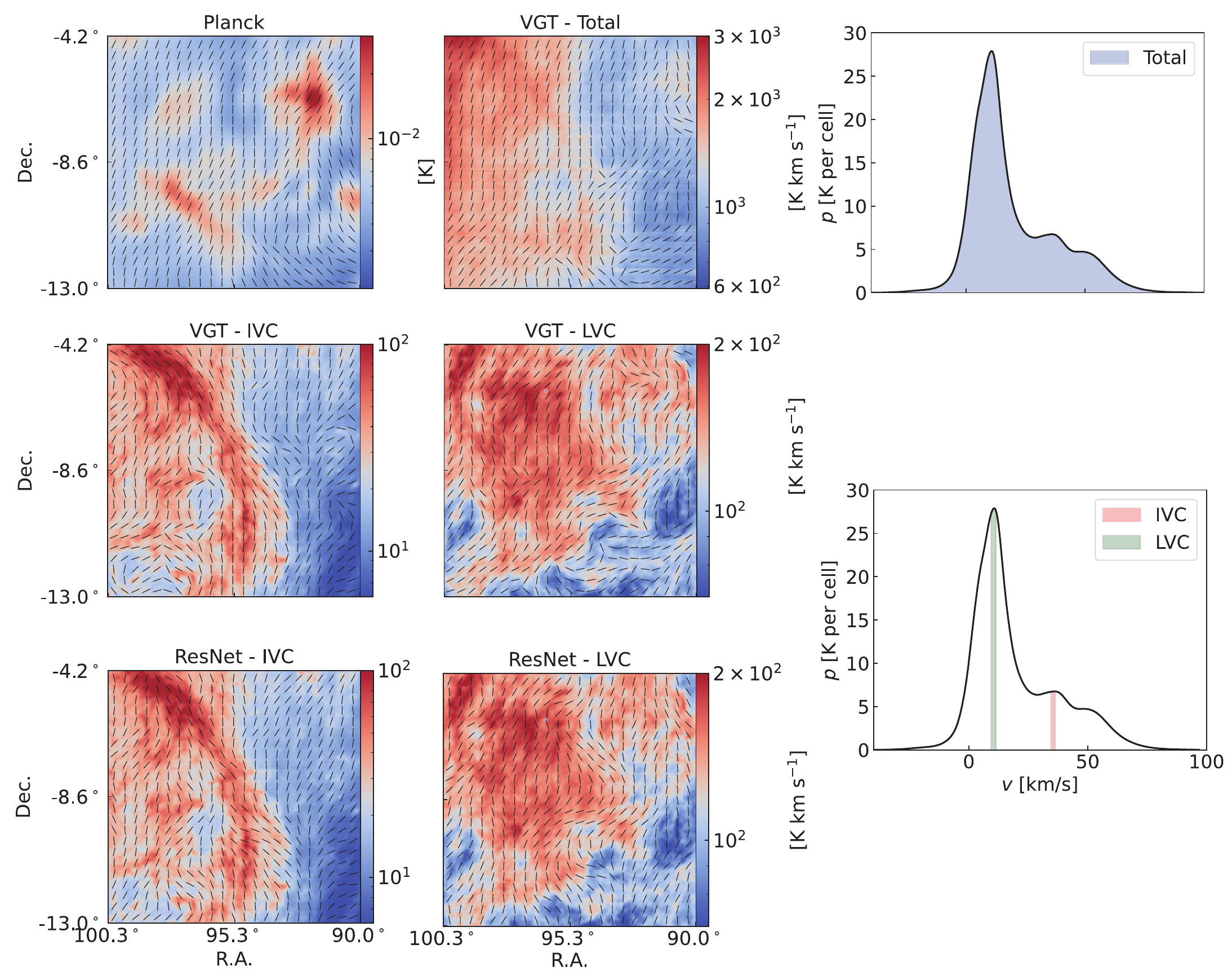}
        \caption{Top: A comparison between the POS magnetic field orientations derived from VGT and those obtained from Planck polarization at 353~GHz. The VGT integrates the contributions from all channels covering the velocity range of [\,-40\,km\,s$^{-1}$, 100\,km\,s$^{-1}$\,], as indicated in the spectrum by blue color. The Planck measurement overlaps with the dust emission intensity map, while the VGT overlaps with the integrated H I intensity map. Middle and bottom: The POS magnetic field orientations traced by the VGT and predicted by the conditional residual neural network (ResNet) for the IVC and LVC, as well as their spectroscopic line. The magnetic fields are overlapped on the spectroscopic channel maps with a width of $2$~km\,s$^{-1}$. The velocities of IVC and LVC are given in the spectrum by red and green colors, respectively.}
    \label{fig:vgt}
\end{figure*}

A direct comparison with the ground truth shows that the predictions for $\psi$, $\gamma$, $B$, and $M_s$ closely resemble their true distributions. However, while the overall $M_A$ distribution is globally similar, its predicted version exhibits numerous localized, spotty features that are absent in the ground truth. This discrepancy in $M_A$ predictions is consistent with the findings of \citet{2024MNRAS.52711240H}, who also reported less accurate predictions of $M_A$ in super-Alfv\'enic conditions using a Convolutional Neural Network architecture. These results suggest that the neural network architecture may not be optimal for capturing the detailed features associated with super-Alfv\'enic turbulence.

\subsection{Uncertainties in the prediction}
\subsubsection{Test with seen data}
Fig.~\ref{fig:sigma} shows the histograms of the uncertainty $\sigma$ in predictions. $\sigma$ is defined as the absolute difference between the predicted value and the ground truth. For data that is already used in the training process, $\sigma_\psi$ and $\sigma_\gamma$ are less than 2 degrees. $\sigma_\psi$ and $\sigma_\gamma$ increase when the magnetic field is relatively weak and velocity dispersion is large, i.e., more super-Alfv\'en conditions. The trend is also true for $\sigma_{M_A}$ and $\sigma_{M_s}$, while they are smaler than 0.2. $\sigma_B$ is smaller than \SI{0.05}{\micro G}, but also slightly increases in weakly magnetized conditions.

\subsubsection{Test with unseen data}
Moreover, we assess the neural network’s performance using a new simulation—not included in the training set—with a mean magnetic field of approximately \SI{4}{\micro G} and a velocity dispersion of about 2.5 km s$^{-1}$. The distributions of $\sigma_\psi$, $\sigma_\gamma$, $\sigma_B$, $\sigma_{M_A}$, and $\sigma_{M_s}$ for this simulation are also shown in Fig.~\ref{fig:sigma}. Although these distributions still peak near zero, they are broader than those derived from the training data. In this simulation, $\sigma_\psi$, $\sigma_\gamma$ reach maximum values near 10$^\circ$ and 15$^\circ$, respectively, with dispersions of roughly 2$^\circ$ and 3$^\circ$. The maximum uncertainty in B is around \SI{1}{\micro G}, with a dispersion of about \SI{0.18}{\micro G}, while $\sigma_{M_A}$ and $\sigma_{M_s}$ can reach approximately 0.4 and 1.0, respectively, with dispersions near 0.08 and 0.16. Additionally, to assess the effects of observational noise and finite beam size, Gaussian noise is added - reducing the signal-to-noise ratio to 20 - and Gaussian blur smoothing is applied to the input image, thereby lowering the resolution by a factor of 2. As illustrated in Fig.~\ref{fig:sigma}, these effects result in negligible changes to the distributions, indicating that the neural network is insensitive to noise and beam size when adequately trained.

\begin{figure*}
\centering
\includegraphics[width=0.8\linewidth]{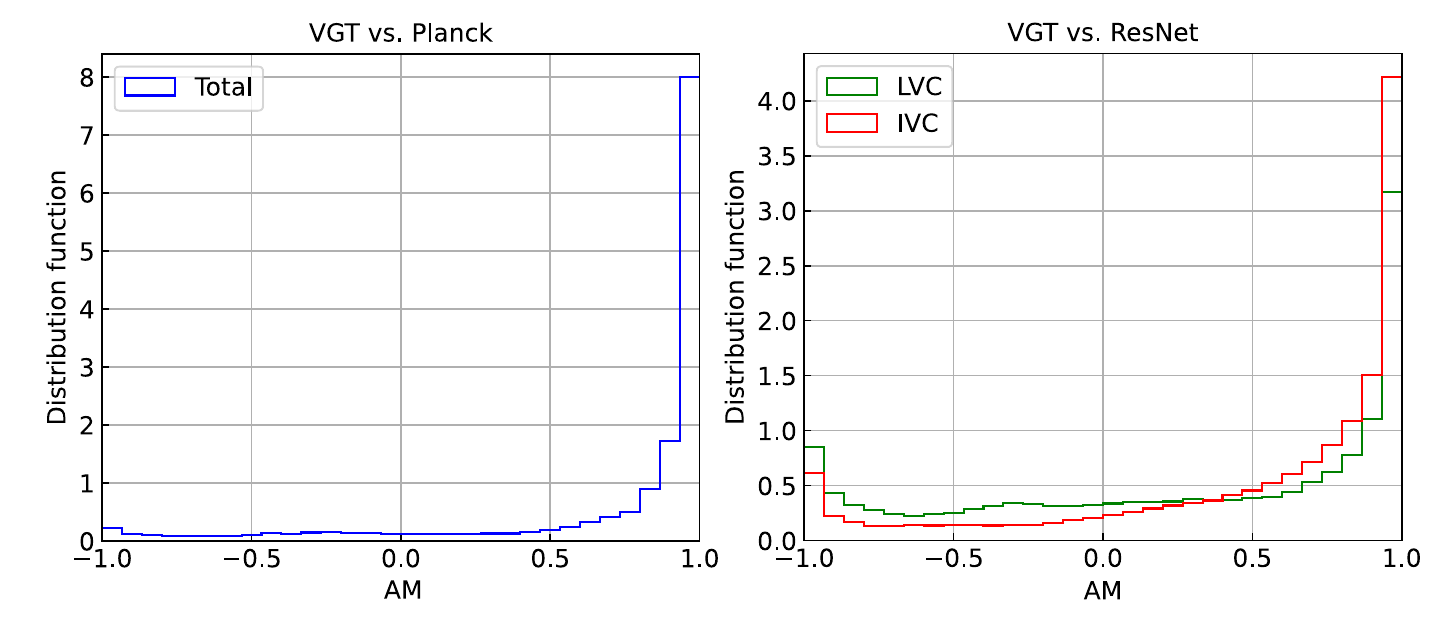}
        \caption{Distribution functions of the AM for VGT vs. Planck (top) and VGT vs. ResNet. A positive AM corresponds to parallel alignment between the two vectors, while a negative AM represents a perpendicular alignment (see Eq.~\ref{eq.am}).}
    \label{fig:am}
\end{figure*}

\subsection{Prediction of the 3D magnetic fields in IVC and LVC}
\subsubsection{Comparison of the POS magnetic field orientation: VGT}
Polarization integrates the signal along the LOS, providing only the integrated POS magnetic field orientation rather than the full 3D distribution. To benchmark the neural network's prediction of the POS magnetic field orientation, we employ the Velocity Gradient Technique (VGT; \citealt{2018ApJ...853...96L,2018MNRAS.480.1333H}). VGT leverages the anisotropy of MHD turbulence imprinted in spectroscopic observations to trace magnetic field structures. As demonstrated by \cite{2020RNAAS...4..105H} through comparisons with stellar polarization, VGT efficiently maps the distribution of the POS magnetic field orientation along the LOS.

Fig.~\ref{fig:vgt} compares the POS magnetic field orientations derived from VGT and those obtained from Planck polarization at 353~GHz \citep{2020A&A...641A...3P}. We applied the VGT pipeline described in \cite{2023MNRAS.524.2379H} to every spectroscopic channel covering the velocity range of [\,-40\,km\,s$^{-1}$, 100\,km\,s$^{-1}$\,] and integrated the contributions from all channels. This integration mimics the polarization signal, under the assumption that H\,I gas, and dust are well mixed. The close correspondence between the orientations derived from VGT and Planck polarization underscores the efficacy of VGT. Moreover, we can discern their distinct magnetic field structures by applying VGT separately to spectroscopic channels corresponding to IVC and LVC. As shown in Fig.~\ref{fig:vgt}, the IVC displays a filamentary morphology, whereas the LVC appears clumpy, each with its own characteristic POS magnetic field orientation.

\begin{figure*}
\centering
\includegraphics[width=0.99\linewidth]{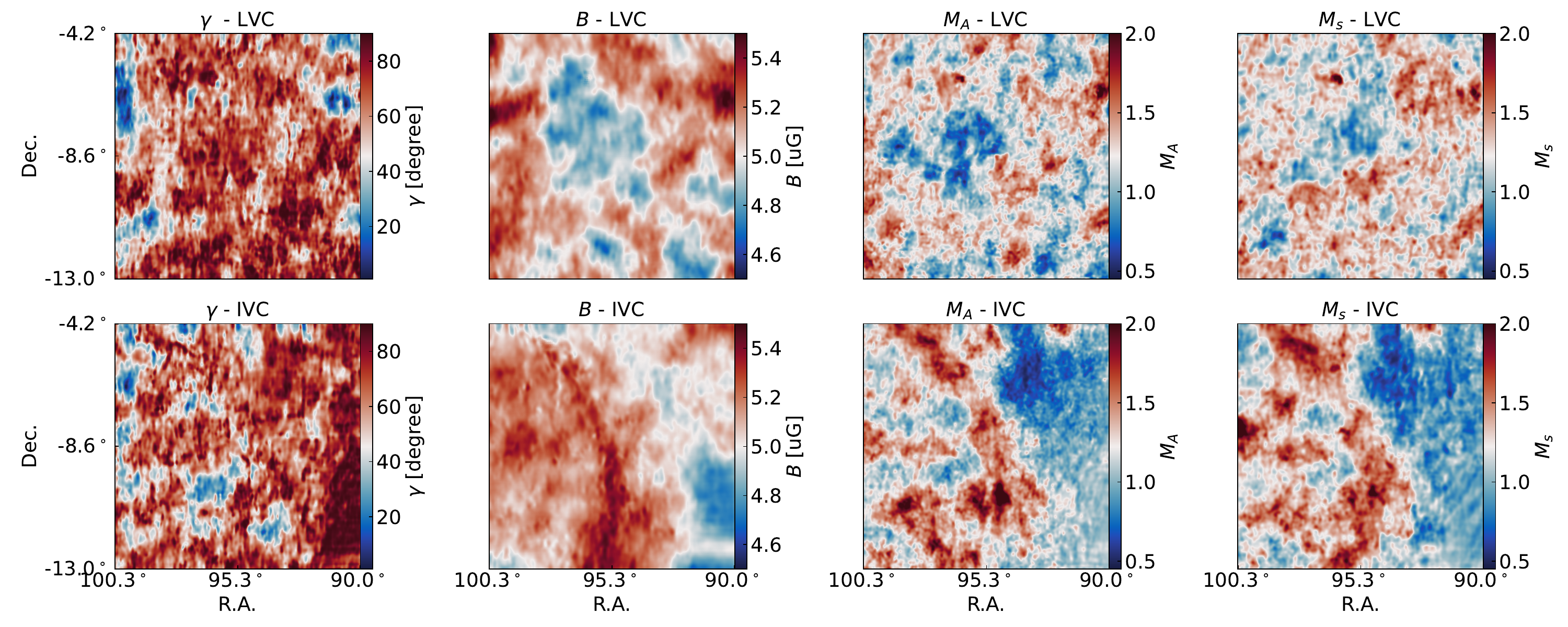}
        \caption{Maps of the predicted $\gamma$, $B$, $M_A$, and $M_s$ for the LVC and IVC (see Fig.~\ref{fig:vgt}).}
    \label{fig:prediction}
\end{figure*}

\subsubsection{Quantification of the agreement}
The POS magnetic field orientation predicted by the residual neural network (i.e., ResNet) is shown in Fig.~\ref{fig:vgt}. To quantify the agreement between the VGT and ResNet, as well as between VGT and Planck polarization measurements, we employ the Alignment Measure (AM) introduced in VGT studies \citep{GL17}. The AM is defined as:
\begin{equation}
\label{eq.am}
\textrm{AM} = 2 \left(\cos^2 \theta_{\rm r} - \frac{1}{2}\right),
\end{equation}
where $\theta_{\rm r}$ represents the relative angle between the two POS magnetic field orientations. An AM value of 1 indicates perfect agreement (i.e., the two are parallel), while an AM of $-1$ denotes perpendicularity.

Fig.~\ref{fig:am} presents the histograms of the AM. When comparing the POS magnetic field orientation integrated along the LOS as traced by VGT and Planck, the AM distribution is sharply peaked around 1, suggesting that VGT accurately traces the magnetic field. When comparing VGT and ResNet for the IVC and LVC, the distributions still concentrate near 1, although the histograms exhibit a spread toward lower AM values. This difference between the ResNet predictions and the VGT results likely arises from uncertainties associated with unseen data and different turbulence conditions (see Fig.~\ref{fig:sigma}).

\subsubsection{3D magnetic fields in IVC and LVC}
Fig.~\ref{fig:prediction} shows the maps of $\gamma$, $B$, $M_A$, and $M_s$ for the LVC and IVC as predicted by ResNet. The two H~I clouds are nearly perpendicular to the LOS, with a median $\gamma$ around 70$^\circ$. The filamentary IVC exhibits a stronger magnetic field (approximately \SI{5.4}{\micro G}) compared to the clumpy LVC, where the field is around \SI{4.8}{\micro G}. In the LVC, the distributions of $M_A$ and $M_s$ appear isotropic, showing no pronounced preferential patterns. Overall, the LVC is characterized by nearly trans-Alfv\'enic ($M_A \approx 1$) and transonic ($M_s \approx 1$) conditions, with the upper limits reaching about 2 for both parameters. 

The IVC shows distinct behavior: its dense filamentary structure is in super-Alfv\'enic conditions with $M_A$ ranging from approximately 1.4 to 1.8, while the surrounding diffuse gas is sub-Alfv\'enic, with $M_A$ between roughly 0.7 and 1.0. Similarly, the filamentary region in the IVC is supersonic with $M_s$ values around 1.5 to 2.0, whereas the ambient medium is subsonic, with $M_s$ between 0.7 and 1.0.

\section{Discussion} 
\label{sec:discussion}
\subsection{Prospects of the deep-learning-based approach}
A comprehensive understanding of the 3D Galactic magnetic field is crucial for addressing many astrophysical problems. These include, but are not limited to, elucidating the physics of star formation \citep{1965QJRAS...6..265M,MK04,MO07,2012ApJ...761..156F,HLS21}, uncovering the origins of ultra-high-energy cosmic rays \citep{2019JCAP...05..004F,2023Sci...382..903T,2025arXiv250215876U}, and accurately modeling Galactic foreground polarization \citep{2014PhRvL.112x1101B,2016A&A...586A.133P}.

Based on the anisotropy of MHD turbulence imprinted in spectroscopic observations, \cite{2024MNRAS.52711240H} proposed the use of a convolutional neural network to trace the 3D magnetic fields in approximately isothermal molecular clouds, deriving the POS position angle, LOS inclination, and Alfv\'en Mach number ($M_A$). In this paper, we extend that deep learning approach to multi-phase H~I and incorporate recent advances in predicting the sonic Mach number ($M_s$; \citealt{2024arXiv241111157S}) and magnetic field strength \citep{2024arXiv241107080Z}. Despite the significant roles that cooling and heating play in multi-phase H~I, we demonstrate that the deep learning approach can predict the full 3D magnetic field (POS position angle, LOS inclination, and strength) as well as the turbulence parameters ($M_A$ and $M_s$).

One of the key advantages of using the H~I emission line is that its velocity information encodes the distance along the LOS. Combined with the Galactic rotational curve, we can ascertain a cloud's spatial position within the Galaxy \citep{2023MNRAS.524.2379H}. Thus, integrating our deep learning approach with the Galactic rotational curve holds the promise of reconstructing the full 3D distribution of 3D magnetic fields within the Galactic disk. Additionally, the deep learning method's application to multiwavelength synchrotron polarization observations, which provides LOS distance information due to the Faraday decorrelation effect, enables the probe of the 3D magnetic field's 3D distribution in the regions of the Galactic halo \citep{2025ApJ...981...58H}.

\subsection{Test the deep-learning approach: anisotropy in young stars' motion}
To test the predictions of our deep‑learning method, we propose exploiting the anisotropy of MHD turbulence as imprinted on the motions of young stars. In magnetized turbulence, eddies become elongated along the local magnetic‐field direction, or equivalently, velocity fluctuations are enhanced perpendicular to the field lines. This distinctive anisotropy encodes both the 3D orientation of the magnetic field and its degree of magnetization (see \S~\ref{sec:theory}). Measuring such anisotropy normally requires full knowledge of the gas distribution's 3D positions and 3D velocity information that is difficult to acquire observationally. However, young stellar associations inherit the turbulent motions of their natal molecular clouds \citep{2021ApJ...907L..40H}, and 6D phase‑space data (3D positions plus 3D velocity) for these stars are now available from the Gaia survey \citep{2023A&A...674A...1G}. By computing the velocity fluctuations of young stars directly from Gaia’s 6D information, one can reconstruct the 3D magnetic‐field geometry and magnetization \citep{2021ApJ...911...37H}, thus providing an independent observational test of the deep‑learning predictions.

\section{Conclusion} 
\label{sec:conclusion}
In conclusion, our study demonstrates that deep learning techniques —specifically, the conditional residual neural network developed in this work — can reconstruct the 3D magnetic fields and turbulence parameters of diffuse H I clouds from spectroscopic observations. The key findings of this research are summarized below:
\begin{enumerate}
    \item By leveraging the anisotropic imprints of MHD turbulence on thin velocity channels and training on synthetic H I data from 3D MHD simulations, our conditional residual neural network simultaneously predicts the POS magnetic field orientation, the LOS inclination, magnetic field strength, sonic Mach number, and Alfv\'en Mach number.
    \item Although the uncertainty in the neural network's predictions increases when processing unseen input data, it remains at a low level overall. Moreover, the model's performance is robust against moderate levels of Gaussian noise and beam smoothing.
    \item Comparisons with established velocity gradient technique confirm that the deep learning predictions for the POS magnetic field orientation are in close agreement.
    \item Application of the neural network to a low-velocity H~I cloud and an intermediate-velocity H~I cloud that overlap in the POS reveals distinct magnetic field structures, strengths, and turbulence characteristics, thereby resolving variations in the clouds' physical conditions along the LOS.
\end{enumerate}

\begin{acknowledgments}
Y.H. acknowledges the support for this work provided by NASA through the NASA Hubble Fellowship grant No. HST-HF2-51557.001 awarded by the Space Telescope Science Institute, which is operated by the Association of Universities for Research in Astronomy, Incorporated, under NASA contract NAS5-26555. This work used SDSC Expanse CPU/GPU and NCSA Delta CPU/GPU through allocations PHY230032, PHY230033, PHY230091, PHY230105,  PHY230178, and PHY240183, from the Advanced Cyberinfrastructure Coordination Ecosystem: Services \& Support (ACCESS) program, which is supported by National Science Foundation grants \#2138259, \#2138286, \#2138307, \#2137603, and \#2138296. 
\end{acknowledgments}

%

\vspace{5mm}

\software{Python3 \citep{10.5555/1593511}
          }



\newpage
\bibliography{sample631}{}
\bibliographystyle{aasjournal}



\end{document}